\documentclass[11pt]{article}
\usepackage{geometry}             
\usepackage{graphicx}
\usepackage{amssymb}
\usepackage{amsmath}
\usepackage{amsthm}
\usepackage{epstopdf}
\usepackage{comment}
\usepackage{cite}
\usepackage{color}
\usepackage[T1]{fontenc}
\usepackage{cancel,tensor,enumitem,fix-cm}
\usepackage{bbold}

\definecolor{brightpink}{rgb}{1.0, 0.0, 0.5}
\usepackage[hyperindex=true,
          pdfstartview=FitH,
          bookmarksnumbered=true,
          bookmarksopen=true,
          citecolor=blue,
          linkcolor=blue,
          colorlinks=true,
          unicode]{hyperref}

\newcommand{\be}{\begin{equation}}
\newcommand{\ee}{\end{equation}}
\newcommand{\bea}{\begin{eqnarray}}
\newcommand{\eea}{\end{eqnarray}}

\usepackage{hyperref}
\hypersetup{
    colorlinks=true,
    linkcolor=black,
    citecolor=black,
    urlcolor=black
}

\begin{document}

\begin{titlepage}
	
	\title{\Large \bf{Energy change and Landauer's principle in the interaction between qubit and quantum field theory}}
	\author{
		Hao Xu
		$$\thanks{{e-mail: }\href{mailto: haoxu@yzu.edu.cn } 
		{ haoxu@yzu.edu.cn} (corresponding author)}
		\vspace{12pt}\\
		\small  $$\textit{Center for Gravitation and Cosmology, College of Physical Science and Technology},\\
		\small \textit{Yangzhou University, Yangzhou, 225009, China}\\
	}
	\date{}
	\maketitle \thispagestyle{empty}

\begin{abstract}
{
We give a general description of the system evolution under the interaction between qubit and quantum field theory up to the second order perturbation, which is also referred to as the simplified model of light-matter interaction. The results are classified into rotating and counter-rotating wave terms, the former corresponding to stimulated absorption and emission, and the latter to Unruh and anti-Unruh effects. We obtain not only the reduced density matrix of the qubit, but also the backreaction obtained by quantum field theory as the environment. The result shows that the energy variation of the quantum field theory is related to trajectory and the initial state of the qubit, the expectation values of the linear and quadratic field operators, and the temporal order product operator. When the qubit is in accelerated motion, the conventional Unruh effect causes the vacuum state to possess a ``temperature'', which raises some doubts about the validity of Landauer's principle. We prove that Landauer's principle still holds for any state of motion.
}

\end{abstract}
\end{titlepage}

\tableofcontents

\section{Introduction}

In modern physics, an isolated or closed system is often described as one that does not exchange energy or information with the external environment and evolves solely through its internal dynamics. However, we know that any isolated system is just an approximation. By definition, a closed system simply can not exist, because if we could not obtain any information about it, we would not be able to study it and confirm its existence. The study of any physical system, large as galaxies or small as elementary particles, is to understand the system through its relationship to all the other physical systems. The collection of all these possible relationships uniquely defines the system, and it is also through these relationships that we, as observers, understand the system. Therefore, the exchange of information between different systems is one of the most important problems in physics.

The way to understand systems through interactions and to study phenomena seen by observers is also the foundation of physics, not only experimentally but also theoretically, and generations of theoretical physicists have constructed thought experiments to help understand physical problems. When trying to study a physical phenomenon or principle, we can start with an ideal detector and then investigate what information it can acquire. The reference in the literature can dates back to the 1980s, when W. G. Unruh and B. DeWitt proposed the so-called Unruh-DeWitt detector model \cite{Unruh:1976db,Unruh:1983ms,DeWitt1979}. Although the Unruh-DeWitt detector is introduced in flat spacetime, the information obtained from an accelerated detector can still help us understand gravity because of the equivalence principle. In curved spacetime, some complicated structures may not admit a time-like Killing vector field, preventing the particle from being defined as the excitation of plane waves in Fock space \cite{Davies}. The particle detectors provide a perspective on this problem from an operational level, i.e. defining a particle as ``what the particle detectors detect'' and reducing to the conventional definition in flat spacetime. Since the observer is present and the system can be studied in quantum open systems \cite{Breuer}, it gives us an intuitive way to study gravity and quantum field theory.

The model of Unruh-DeWitt detector consists of a detector, usually a qubit with two energy levels, and a quantized field. The interaction Hamiltonian $\hat{H}_{\text{int}}\propto (\hat{\sigma}_{-} +\hat{\sigma}_{+})(\hat{a}+\hat{a}^{\dagger})$, where $\hat{\sigma}_{+}=|e\rangle\langle g|$ raises and $\hat{\sigma}_{-}=|g\rangle\langle e|$ lowers the energy of the qubit, and $\hat{a}$ and $\hat{a}^{\dagger}$ are the annihilation and creation operators of the field, respectively. This is also the simplified model of the light-matter interaction. The $\hat{H}_{\text{int}}$ contains four terms, corresponding to four different physical phenomena. When the detector moves inertially in space and is weakly coupled to the field, conservation of energy results in the energy of the absorbed or emitted photon being equal to the energy gap of the detector, corresponding to the \emph{rotating wave terms} $  \hat{\sigma}_{+}\hat{a}+\hat{\sigma}_{-} \hat{a}^{\dagger}$, leading to the so-called Jaynes-Cummings model \cite{Jaynes1963}, which is widely used in atomic physics and quantum optics \cite{Allen1987,Tannoudji2004}. However, if the detector moves non-inertially in free space, regardless of the source providing this motion, the detector and the quantum field no longer satisfy conservation of energy. The physical phenomena corresponds to \emph{counter-rotating wave terms}, i.e. $\hat{\sigma}_{+} \hat{a}^{\dagger}+\hat{\sigma}_{-}\hat{a}$. The traditional statement of the Unruh effect is that the observer accelerating uniformly through empty space will perceive a thermal bath whose temperature is proportional to the acceleration. This is because the $\hat{\sigma}_{+} \hat{a}^{\dagger}$ term allows the qubit and the quantum field theory to increase the energy together, so the quantum field theory may appear to the qubit as a thermal field. Similarly, the $\hat{\sigma}_{-}\hat{a}$ will cause the qubit and the quantum field to lose energy together, if both of them are not in the ground state. We can therefore refer to the counter-rotating wave terms as Unruh and anti-Unruh terms.

The conventional Unruh effect has been studied in many different scenarios, including different spacetimes, states of motion, and definitions of vacuum \cite{Fukuma:2013uxa,Rabochaya:2015aza,Ng:2016hzn,Hotta:2020pmq,Arias:2017kos,Gray:2018ifq,Xu:2019hea,Ng:2018drz,Ahmed:2020fai,Ahmed:2023uem,Pitelli:2021oil,Hodgkinson:2014iua,Xu:2021buk,Polo-Gomez:2023gaz,Perche:2023lwo,Alonso-Serrano:2024sop}. For most cases, the concern remains with the interaction of a qubit in the ground state with the vacuum state of the quantum field theory through some fixed acceleration. However, the crux of the Unruh effect is actually counter-rotating wave terms, and we do not need to limit the discussion to fixed accelerations or vacuum states. Arbitrary non-inertial motions, different quantum field states, and backreaction to the quantum fields are also worth investigating in their own right. It has been recently shown that by a suitable choice of states of motion and quantum field theory, although not in uniformly accelerated motion and vacuum state, we can amplify the observational effects of counter-rotating wave terms, thus making it possible to experimentally demonstrating the existence of the Unruh effect with the current technology \cite{Kempf2022}.

In this work we give a general description of the system evolution under the interaction between qubit and free massless scalar quantum field theory. We will not specify a particular motion state for the detector, nor will we restrict the quantum field theory to the vacuum state. In particular, in addition to the response received by the detector, we also discuss the energy change of the quantum field theory itself during the interaction, an issue that has been largely overlooked in the literature. Previous studies of similar interactions have always made assumptions due to differences in research focus. For example, when studying detectors, we have often assumed that the system being measured is so large that it is not changed by the measurement. On the other hand, when studying variations of quantum field theory, such as in the study of thermalization of conformal field theory \cite{Cardy:2014rqa,Liu:2013iza,Liu:2013qca,Xu:2017wvu}, we often introduce quantum quench by hand in order to test its effect on the field theory, where the quenches are such that the sudden injection of energy density either locally or globally. However, these are all assumptions, and they all focus on only one of the interacting systems, ignoring the reaction of the others. The exchange of information should be reflected in all the interacting systems. Therefore, a general description of this interaction has important implications for our further understanding of both the observation and the reversibility in physical processes.

Furthermore, since the traditional description of the Unruh effect associates a temperature with the vacuum state, this may have led to a confusing definition of temperature, especially in statistical physics and quantum information, where temperature has a unique meaning in certain physical processes. For example, in Landauer's principle, temperature is considered a necessary condition for its validity \cite{landauer1961,Reeb2013}, but the definition of temperature in the Unruh effect seems to become observer-dependent, which may cause some confusion. Since the temperature can vary from observer to observer, one might ask whether the Landauer's principle is also observer dependent. We will prove that although the vacuum in the Unruh effect appears to be ``thermal'', the counter-rotating wave terms do not affect the validity of Landauer's principle.

The remainder of the paper is organized as follows. In section \ref{section2} we study the interaction between the qubit and the quantum field theory by perturbation, calculate the variation of the qubit and the quantum field theory up to the second order, and classify the contributions of the rotating and counter-rotating wave terms. In particular, we also obtain the energy change of the quantum field theory during this process. In section \ref{section3} we investigate the von Neumann entropy of the qubit and the energy of the quantum field theory in terms of the rotating and counter-rotating wave terms, respectively. We find that both of them must satisfy Landauer's principle, although their physical mechanisms are different. In section \ref{section4} we summarize the results and suggest some possible directions for future research. We set the speed of light in vacuum, the gravitational constant, the Planck constant and the Boltzmann constant all equal to unity.

\section{The model}
\label{section2}
In this section we give a general description of the interaction between a qubit and quantum field theory, which is often referred to in the literature as the light-matter interaction. The total Hamiltonian $\hat{H}_{\text{total}}$ of the system can be written as
\begin{equation}
\hat{H}_{\text{total}}=\hat{H}_0+\hat{H}_{\text{int}},
\end{equation}
and the $\hat{H}_0$ is the sum of the free Hamiltonian of the qubit and the quantum field theory
\begin{equation}
\hat{H}_0=\hat{H}+\hat{H}_{\text{field}},
\end{equation}
where  $\hat{H}=\frac{\Omega}{2}{{\hat{\sigma}}_z}$, $\hat{\sigma}_z$ is the Pauli matrix, and $\Omega$ is the energy gap between ground state $|g\rangle $ and excited state $|e \rangle $ in the rest frame. For simplicity we choose the quantum field to be free massless scalar field. The Hamiltonian $\hat{H}_{\text{field}}$ in free space is usually written as a particle number operator multiplied by the frequency and then integrated over momentum space. However, in this work we discretize the quantum field theory and choose a cutoff such that the $\hat{H}_{\text{field}}=\sum_{j=1}^{l}\omega_j \hat{a}^{\dag}_j\hat{a}_j$, where $l$ corresponds to the largest momentum. This model can be used to simulate how atoms interact with electromagnetic fields. If we turn the summation into an integral and let $l$ tend to infinity, we can obtain the quantum field theory in free space.

The interaction Hamiltonian is given by
\begin{equation}
\hat{H}_{\text{int}}=\lambda \hat{\sigma}_x\hat{\phi}[x(\tau)],
\end{equation}
where $\lambda$ is a weak coupling constant, $\hat{\sigma}_x$ is monopole operator which allows population to be exchanged between energy levels, $\tau$ denotes proper time, and $\hat{\phi}[x(\tau)]$ is the field operator at the position of the qubit. In the interaction picture, we have
\begin{equation}
\hat{\sigma}_x(\tau)=\hat{\sigma}^{+}e^{i\Omega \tau}+\hat{\sigma}^{-}e^{-i\Omega \tau},
\end{equation}
and
\begin{align}
\hat{\phi}[x(\tau)] = \sum_{j=1}^{l}\left( \hat{a}_je^{-i\omega_j t(\tau)+i\mathbf{k_j \cdot x(\tau)}}+\hat{a}^{\dagger}_je^{i\omega_j t(\tau)-i\mathbf{k_j \cdot x(\tau)}} \right).
\label{int}
\end{align}

 The time evolution operator of the system from time $\tau=0$ to $\tau=s$ is given by the Dyson series:
\begin{align}
\hat{U}(s,0)=  &\mathbb{1} \underbrace{-i\int^{s}_{0}d\tau \hat{H}_{\text{int}}(\tau)}_{\hat{U}^{(1)}} \nonumber \\ 
&\underbrace{+(-i)^2\int^{s}_{0}d\tau_1 \int^{\tau_1}_{0}d\tau_2 \hat{H}_{\text{int}}(\tau_1)\hat{H}_{\text{int}}(\tau_2)}_{\hat{U}^{(2)}}+ ...\nonumber \\ 
&\underbrace{+(-i)^n\int^{s}_{0}d\tau_1 ... \int^{\tau_{n-1}}_{0}d\tau_{n} \hat{H}_{\text{int}}(\tau_1) ... \hat{H}_{\text{int}}(\tau_{n})}_{\hat{U}^{(n)}},
\label{dyson}
\end{align}
so the density matrix of the total system can be written as
\begin{equation}
\rho=\rho^{(0)}+\rho^{(1)}+\rho^{(2)}+\mathcal{O}(\lambda^3),
\end{equation}
where
\begin{align}
\rho^{(0)}&=\rho_0, \\
\rho^{(1)}&=\hat{U}^{(1)}\rho_0+\rho_0 \hat{U}^{(1)\dagger}, \\
\rho^{(2)}&=\hat{U}^{(1)}\rho_0 \hat{U}^{(1)\dagger}+\hat{U}^{(2)}\rho_0+\rho_0 \hat{U}^{(2)\dagger}.
\label{rho}
\end{align}

We choose the initial state of the qubit to be 
\begin{equation}
\rho_{\text{0}}^q=\begin{pmatrix}
    p & 0 \\

    0 & 1-p
\end{pmatrix},
\label{dq}
\end{equation}
where $0< p< 1$ is a real number. In principle, the density matrix can also contain off-diagonal elements, thus allowing us to study the effect of decoherence. However, decoherence is not the focus of this study, otherwise we would still need to diagonalize the density matrix when calculating the von Neumann entropy, and we can always choose a set of orthogonal normalized basis vectors such that the density matrix of the qubit is already diagonal and thus the decoherence effect can be ignored. See e.g. \cite{Nesterov:2020exl,Xu:2022juv,Xu:2023tdt,Danielson:2022tdw,Danielson:2022sga,Danielson:2024yru,Wilson-Gerow:2024ljx,Biggs:2024dgp} for recent discussions on decoherence.

We choose the initial state for the quantum field theory to be
\begin{equation}
\rho_{\text{0}}^{f}= \bigotimes_{j=1}^{l} \rho_j,
\end{equation}
where the $\rho_j$ is the density matrix of the modes $k_j$, and we do not place any restrictions on the form of $\rho_j$. The inital state of the total system is then 
\begin{equation}
\rho_{0}=\begin{pmatrix}
    p & 0 \\

    0 & 1-p
\end{pmatrix} \otimes \bigotimes_{j=1}^{l} \rho_j.
\end{equation} 
Now we can solve the time evolution of the total system order by order.

\subsection{The order of $\lambda$}

The operator $\hat{U}^{(1)}$ can be written as
\begin{equation}
\hat{U}^{(1)} = \frac{\lambda}{i}\sum_{j=1}^{l}(\hat{\sigma}^+\hat{a}_j^{\dagger}I_{+,j}+\hat{\sigma}^{+} \hat{a}_j I_{-,j}^{*}+\hat{\sigma}^{-}\hat{a}_j^{\dagger}I_{-,j}+ \hat{\sigma}^{-} \hat{a}_j I_{+,j}^{*}),
\end{equation}
where
\begin{equation}
I_{\pm,j}:=\int^{s}_0 d\tau~e^{i\left[\pm \Omega \tau+\omega_jt(\tau)-\mathbf{k_j \cdot x(\tau)}\right]}.
\label{I}
\end{equation}
We can easily find that $I_{-,j}$ and $I_{-,j}^{*}$ correspond to the rotating wave terms, while $I_{+,j}$ and $I_{+,j}^*$ correspond to the counter-rotating wave terms. In the inertial motion, the trajectory of the qubit is given by $x^{\mu} (\tau) = (\gamma \tau, \gamma \mathbf{v}\tau)$ and $\gamma=(1-|{\mathbf{v}}|^2)^{-1/2}$, and we have
\begin{equation}\label{inertialA}
I_{\pm,j}\propto \delta\left(\Omega \pm \gamma(\omega_{j} - \mathbf{k_j}\cdot \mathbf{v}) \right),
\end{equation}
which is non-zero only on resonance. Only if the energy gap of the qubit matches the relativistically
Doppler-shifted energy of the field, the $I_{-,j}$ contributes, while $I_{+,j}$ is identically zero in all cases because $\left(\omega_{j} - \mathbf{k_j}\cdot \mathbf{v}\right)$ is always non-negative.

In the non-inertial motion, however, the above conclusion would no longer hold. For example, in uniformly accelerated motion, the trajectory is given by $x^{\mu} (\tau) = (\frac{1}{a}\sinh(a\tau), \frac{1}{a}\cosh(a\tau))$, where $a$ denotes the acceleration. The transition rate of the qubit is proportional to the $|I_{+,j}|^2$, where the integral is over the entire time axis. We have
\begin{equation}
|I_{+,j}|^2 \propto \frac{1}{e^{\frac{2\pi \Omega}{a}}-1},
\end{equation}
and $|I_{-,j}|^2 = e^{\frac{2\pi \Omega}{a}}|I_{+,j}|^2$. This is the conventional Unruh effect \cite{Ben-Benjamin:2019opz}.

In most research, the rotating and the counter-rotating wave terms are considered separately. In atomic physics or quantum optics, the focus is often the stimulated absorption and emission, so the rotating wave approximation is usually considered as a prerequisite in the Jaynes-Cummings model or using the Lindblad master equation to study open systems. On the other hand, in the study of the Unruh effect, one usually considers the interaction between an accelerated qubit in the ground state and the quantum field theory in the vacuum state, and thus is concerned only with the contribution of counter-rotating wave terms, precisely on $\sigma_+a^{\dagger}$ alone. In this work, we do not impose any constraints on the state of motion of the qubit or the state of the quantum field, which allows us to discuss cases that are as universal as possible.

We can calculate the $\hat{U}^{(1)}\rho_0+\rho_0 \hat{U}^{(1)\dagger}$ directly and do the partial trace of the quantum field theory, so we have the reduced density matrix of the qubit written as
\begin{equation}
\rho_{\text{0}}^q=\begin{pmatrix}
    0 & \delta d \\

   \delta d^* & 0
\end{pmatrix},
\label{dq}
\end{equation}
where
\begin{equation}
\delta d= i\lambda \sum_{j=1}^{l}(2p-1)\left( \langle \hat{a}_j\rangle I_{-,j}^{*} + \langle \hat{a}_j^{\dagger}\rangle I_{+,j}\right).
\label{one}
\end{equation}
This is the first main result of this work, and $\delta d$ corresponds to the one-point function of quantum field theory. If the field is in the vacuum state, or if the density matrix $\rho_j$ are all diagonal, we have $\langle a_j^{(\dagger)}\rangle=0$, i.e. the first order contribution is zero, which is why much of the literature studies the conventional Unruh effect from second order. On the other hand, since the density matrix of the qubit does not contain diagonal elements, doing a partial trace on the qubit only yields zero, which means that the quantum field theory does not change at this order.

\subsection{The order of $\lambda^2$}

First we calculate $\hat{U}^{(1)}\rho_0 \hat{U}^{(1)\dagger}$ directly, which gives us
\begin{equation}
\hat{U}^{(1)}\rho_0 \hat{U}^{(1)\dagger}=\lambda^2\iint d \tau_1 d\tau_2 \begin{pmatrix}
    (1-p)e^{-i\Omega(\tau_1-\tau_2)} & 0 \\

    0 & pe^{i\Omega(\tau_1-\tau_2)}
\end{pmatrix} \hat{\phi}(\tau_2)\bigotimes_{j=1}^{l} \rho_j \hat{\phi}(\tau_1)
\end{equation}
Taking the partial trace of the quantum field theory, we can have the reduced density matrix of the qubit written as
\begin{equation}
\lambda^2\iint  d \tau_1 d\tau_2 \begin{pmatrix}
    (1-p)e^{-i\Omega(\tau_1-\tau_2)} & 0 \\

    0 & pe^{i\Omega(\tau_1-\tau_2)}
\end{pmatrix} \langle \hat{\phi}(\tau_1) \hat{\phi}(\tau_2)\rangle,
\end{equation}
where $\langle \hat{\phi}(\tau_1) \hat{\phi}(\tau_2)\rangle$ is the two-point function of the quantum field theory. Similarly we can also calculate $\hat{U}^{(2)}\rho_0+\rho_0 \hat{U}^{(2)\dagger}$, but it is worth noting that temporal order product operator is required in $\hat{U}^{(2)}$. We have
\begin{equation}
\hat{U}^{(2)}\rho_0 =- \theta(\tau_1-\tau_2) \lambda^2 \iint d \tau_1 d\tau_2 \begin{pmatrix}
    pe^{i\Omega(\tau_1-\tau_2)} & 0 \\

    0 & (1-p)e^{-i\Omega(\tau_1-\tau_2)}
\end{pmatrix} \hat{\phi}(\tau_1) \hat{\phi}(\tau_2) \bigotimes_{j=1}^{l} \rho_j
\end{equation}
and 
\begin{equation}
\rho_0 \hat{U}^{(2)\dagger} =- \theta(\tau_2-\tau_1) \lambda^2 \iint d \tau_1 d\tau_2 \begin{pmatrix}
    pe^{i\Omega(\tau_1-\tau_2)} & 0 \\

    0 & (1-p)e^{-i\Omega(\tau_1-\tau_2)}
\end{pmatrix} \bigotimes_{j=1}^{l} \rho_j \hat{\phi}(\tau_1) \hat{\phi}(\tau_2).
\end{equation}

Fortunately, when we do the partial trace of the quantum field theory, both $\hat{U}^{(2)}\rho_0$ and $\rho_0 \hat{U}^{(2)\dagger}$ still correspond to the two-point function and are identical except for the step function $\theta$. Since $\theta(\tau_1-\tau_2) + \theta(\tau_2-\tau_1) = 1$, we know that the reduced density matrix of the qubit obtained for $\hat{U}^{(2)}\rho_0+\rho_0 \hat{U}^{(2)\dagger}$ is
\begin{equation}
- \lambda^2 \iint d \tau_1 d\tau_2 \begin{pmatrix}
    pe^{i\Omega(\tau_1-\tau_2)} & 0 \\

    0 & (1-p)e^{-i\Omega(\tau_1-\tau_2)}
\end{pmatrix}  \langle\hat{\phi}(\tau_1) \hat{\phi}(\tau_2)\rangle.
\end{equation}
Therefore, the reduced density matrix of qubit at second order can be written as
\begin{equation}
\begin{pmatrix}
    \delta p & 0 \\

    0 & -\delta p
\end{pmatrix},
\end{equation}
where
\begin{equation}
\delta p= \lambda^2 \iint d \tau_1 d\tau_2 \left( (1-p)e^{-i\Omega(\tau_1-\tau_2)}-pe^{i\Omega(\tau_1-\tau_2)}\right) \langle\hat{\phi}(\tau_1) \hat{\phi}(\tau_2)\rangle .
\label{p}
\end{equation}

Now we give a brief discussion of the two-point function. Defining $u_j(x):=e^{-i\omega_j t(\tau)+i\mathbf{k_j \cdot x(\tau)}}$, the two-point function can be written as
\begin{equation}
\langle\hat{\phi}(\tau_1) \hat{\phi}(\tau_2)\rangle = \left\langle \sum_{m=1}^{l}(\hat{a}_m u_m(x_1)+ \hat{a}^{\dagger}_m u_m^*(x_1) )\sum_{n=1}^{l}(\hat{a}_n u_n(x_2)+ \hat{a}^{\dagger}_n u_n^*(x_2) ) \right\rangle.
\end{equation}

In traditional quantum field theory, we often discuss correlation functions in vacuum states. For calculations in non-vacuum states, such as S-matrices, we also use methods like the LSZ reduction formula to convert the problems to the vacuum state. However, this approach is not always successful, since states can in principle be arbitrary and even contain an infinite number of creation operators. When written in terms of the expectation values of the creation and annihilation operators, the two-point function contains a total of $4l^2$ terms. 

If the density matrix of the quantum field theory is diagonal, all the $m\neq n$ terms, i.e., the linear terms of $ \langle \hat{a} \rangle$ and $ \langle \hat{a}^{\dagger} \rangle$, are zero, and both $ \langle \hat{a}^2 \rangle$ and $ \langle \hat{a}^{\dagger2} \rangle$ in $m=n$ are also zero. When the quantum field theory is in the vacuum state, $\langle \hat{a}^{\dagger}\hat{a} \rangle$ vanishes too, and only $\langle \hat{a} \hat{a}^{\dagger} \rangle$ remains due to the canonical commutation relation. Thus in the textbook the correlation function for the vacuum state is usually directly expressed as $\int dk u_k(x_1)u^*_k(x_2)$. It corresponds to the quantum effects of field theory and is present for any quantum field state.

For general states, however, we need to discuss the cases $m \neq n$ and $m = n$ separately, which correspond to the expectation values of the linear and quadratic terms of the operator, respectively. By classifying the terms in the two-point function, we have
\begin{align}
\langle\hat{\phi}(\tau_1) \hat{\phi}(\tau_2)\rangle &= \sum_{m=1}^l\sum_{n(\neq m)=1}^l \big[\langle \hat{a}_m \rangle \langle \hat{a}_n \rangle u_m(x_1)u_n(x_2) + \langle \hat{a}_m \rangle \langle \hat{a}^{\dagger}_n \rangle u_m(x_1)u^*_n(x_2) \nonumber \\
&+ \langle \hat{a}^{\dagger}_m \rangle \langle \hat{a}_n \rangle u^*_m(x_1)u_n(x_2) + \langle \hat{a}^{\dagger}_m \rangle \langle \hat{a}^{\dagger}_n \rangle u^*_m(x_1)u^*_n(x_2) \big]\nonumber \\
&+\sum_{j=1}^l\big[  \langle \hat{a}_j^2 \rangle u_j(x_1)u_j(x_2) + \langle \hat{a}_j \hat{a}^{\dagger}_j \rangle u_j(x_1)u^*_j(x_2) +
\langle \hat{a}^{\dagger}_j \hat{a}_j \rangle u^*_j(x_1)u_j(x_2)\nonumber \\
&+ \langle {\hat{a}_j^{\dagger2}} \rangle u^*_j(x_1)u^*_j(x_2) \big].
\label{two}
\end{align}
In the first summations, $m\neq n$ means that the field operators in the two-point function are not being in same momentum modes, while in the second summation over $j$, the field operators are in the same momentum. After obtaining the two-point function, we can insert it into the eq.\eqref{p} and integrate over time to obtain the final state of the qubit.

However, energy change of the quantum field theory could be much more complicated. This is because that after doing the partial trace of the qubit, the correction to the field part of $\hat{U}^{(1)}\rho_0 \hat{U}^{(1)\dagger}$ corresponds to $\hat{\phi}(\tau_2)\bigotimes_{j=1}^{l} \rho_j \hat{\phi}(\tau_1)$, while $\hat{U}^{(2)}\rho_0$ and $\rho_0 \hat{U}^{(2)\dagger}$ correspond to $- \theta(\tau_1-\tau_2)\hat{\phi}(\tau_1) \hat{\phi}(\tau_2) \bigotimes_{j=1}^{l} \rho_j$ and $- \theta(\tau_2-\tau_1)\bigotimes_{j=1}^{l} \rho_j \hat{\phi}(\tau_1) \hat{\phi}(\tau_2)$, respectively. In order to calculate the energy, we need to act the $\hat{H}_{\text{field}}=\sum_{j=1}^{l}\omega_j \hat{a}^{\dag}_j\hat{a}_j$ on the above states and solve for trace, which would make each term correspond to a $4l^3$ subterms, and many of them are not commute with each other, and the temporal order product operator do not always cancel out. 

Fortunately, through a series of complicated calculations, we have the change of energy as
\begin{align}
\frac{\Delta E}{\lambda^2}&= p \sum_{n(\neq m)=1}^l \omega_n \langle \hat{a}^{\dagger}_n \rangle I_{-,n}(\tau_2) \sum_{m=1}^l \langle \hat{a}_m \rangle I^*_{-,m}(\tau_1)+(1-p) \sum_{n(\neq m)=1}^l \omega_n \langle \hat{a}^{\dagger}_n \rangle I_{+,n}(\tau_2) \sum_{m=1}^l \langle \hat{a}_m \rangle I^*_{+,m}(\tau_1)  \nonumber \\
&- p \sum_{n(\neq m)=1}^l \omega_n \langle \hat{a}_n \rangle I^*_{+,n}(\tau_2) \sum_{m=1}^l \langle \hat{a}^{\dagger}_m \rangle I_{+,m}(\tau_1)-(1-p) \sum_{n(\neq m)=1}^l \omega_n \langle \hat{a}_n \rangle I^*_{-,n}(\tau_2) \sum_{m=1}^l \langle \hat{a}^{\dagger}_m \rangle I_{-,m}(\tau_1)\nonumber \\
&+\varepsilon(\tau_1-\tau_2)(4p-2)\text{Re}\left(\sum_{m(\neq n)=1}^l \omega_m \langle \hat{a}_m \rangle I^*_{-,m}(\tau_1)\sum_{n=1}^l \langle \hat{a}_n \rangle I^*_{+,n}(\tau_2)\right) \nonumber \\
&+\sum_{j=1}^l \omega_j \left[\left(p\langle \hat{a}_j \hat{a}^{\dagger}_j \rangle- (1-p)\langle \hat{a}^{\dagger}_j \hat{a}_j \rangle\right)|I_{-,j}|^2-\left(p\langle \hat{a}^{\dagger}_j \hat{a}_j \rangle -(1-p)\langle \hat{a}_j \hat{a}^{\dagger}_j \rangle\right)|I_{+,j}|^2\right] \nonumber \\
&+\varepsilon(\tau_1-\tau_2)(4p-2)\text{Re}\left(\sum_{j=1}^l \omega_l \langle \hat{a}^2_j \rangle I^*_{-,j}(\tau_1) I^*_{+,j}(\tau_2)\right)
\label{E}
\end{align}
We can find the energy change of the quantum field theory is related to trajectory and the initial state of the qubit, the expectation values of the linear and quadratic field operators, and the temporal order product operator. The eq.\eqref{two} and \eqref{E} are the second main result of this work, and together with eq.\eqref{one} they constitute the variation of the density matrix of the qubit and the energy of the field theory up to the second order. 

For almost all states, the two-point function and energy changes of the field theory need to be calculated in the way as eq.\eqref{two} and \eqref{E}, where the linear and quadratic expectation values of the operator are independent. However, in the coherent field, the state is the eigenstate of $\hat{a}$, so the quadratic order can also be expressed directly in terms of the linear order and we do not need to categorize any more. Since the two-point function can be expressed in terms of one-point functions plus the vacuum part, the effect of coherent cancels out in the diagonalization of the qubit, leaving only the contribution of the vacuum state. This also allows a pair of detectors to harvest the same amount of entanglement from any coherent field state as from the vacuum \cite{Simidzija:2017jpo,Simidzija:2017kty}. 

Higher order perturbations can be solved in the same way, but the calculation becomes more difficult as the order increases. If the quantum field theory is a Gaussian state, the expectation values of operators in higher order perturbations are fully described by the first and second moments of their quadratures \cite{Adesso2007,Adesso2014}, so they can be obtained from the results of $\lambda$ and $\lambda^2$ order pertubation.

\section{Validity of Landauer's principle}
\label{section3}

Since most ``environments'' can be approximated as a thermal state, in this section we consider the quantum field theory in the thermal state. For our model, the expectation value of linear terms of the operator is zero in thermal state, while the particle number is written in Bose-Einstein form $\langle \hat{a}^{\dagger}_j \hat{a}_j \rangle:=\bar{n}_j={1}/{\left(e^{\frac{\omega_j}{T}}-1\right)}$. This makes the calculation of eq. \eqref{one}, \eqref{two} and \eqref{E} to be greatly simplified. 

Moreover, the thermal state is also a condition for Landauer's principle to hold. Landauer's principle relates the entropy change of a system to the heat dissipated into the environment during any logically irreversible computation, providing a theoretical limit of energy consumption throughout the process \cite{landauer1961,Reeb2013}. If the following four conditions are met: (i) both the system and environment are quantum, (ii) environment is initially thermal, (iii) system and environment are initially uncorrelated, and (iv) the process proceeds by unitary evolution, then the Landauer's principle can be expressed as 
\begin{equation}
\Delta Q\geqslant {\color{red}}T \Delta S,
\label{bound}
\end{equation}
where $\Delta Q=Q_f-Q_{0}$ and $T$ are the energy change and the temperature the environment, respectively. The $\Delta S=S_{0}-S_{f}$ is the difference between the initial and final state von Neumann entropy of the system. The subscripts $f$ and $0$ stand for final and initial state, respectively. We can set the qubit as the system and the quantum field theory as the environment.

In the previous work \cite{Xu:2024xlx}, it was found that in the rotating wave approximation, Landauer's principle and temperature are closely related, and the Bose-Einstein distribution is a natural corollary. However, for counter-rotating wave terms, even the vacuum state can be associated with a “temperature” by the Unruh effect, which casts some doubt on the validity of Landauer's principle. Since different observers can have different opinions about the temperature, previous results in the rotating wave terms may no longer be applicable, and one may wonder if the validity of Landauer's principle also depends on the observer. In this section we will show that Landauer's principle still holds for counter-rotating wave terms.

Directly calculating the two-point function eq.\eqref{two} and inserting it into eq.\eqref{p}, we have
\begin{equation}
\delta p= \lambda^2\sum_{j=1}^{l} \left(-[(\bar{n}_j+1)p-\bar{n}_j(1-p)]|I_{-,j}|^2 +[(\bar{n}_j+1)(1-p)-\bar{n}_jp]|I_{+,j}|^2\right),
\end{equation}
and
\begin{equation}
\Delta S = \lambda^2\sum_{j=1}^{l}\ln{\frac{1-p}{p}}\Big\{[(\bar{n}_j+1)p-\bar{n}_j(1-p)]|I_{-,j}|^2 -[(\bar{n}_j+1)(1-p)-\bar{n}_jp]|I_{+,j}|^2\Big\}.
\end{equation}

Similarly, we can calculate the energy change in the quantum field theory to get
\begin{equation}
\frac{\Delta Q}{T}= \lambda^2\sum_{j=1}^{l}\ln{\frac{\bar{n}_j+1}{\bar{n}_j}}\Big\{[(\bar{n}_j+1)p-\bar{n}_j(1-p)]|I_{-,j}|^2 +[(\bar{n}_j+1)(1-p)-\bar{n}_jp]|I_{+,j}|^2\Big\}.
\end{equation}

When $0<p<1/2$, the qubit in $(1-p)|g\rangle\langle g|+p|e\rangle\langle e|$ corresponds to an effective temperature $T_q$ satisfying $p=1/(e^{\frac{\Omega}{T_q}}+1)$. Assuming $|I_{+,j}|^2 = 0$, as in inertial motion, the change of the total system comes from the rotating wave terms, corresponding to stimulated absorption and emission. The $\Delta S$ and ${\Delta Q}/{T}$ share the same sign, depending on the values of $\frac{\bar{n}_j+1}{\bar{n}_j}$ and $\frac{1-p}{p}$. If $\frac{\bar{n}_j+1}{\bar{n}_j}>\frac{1-p}{p}$, the energy of the qubit is transferred to the quantum field theory, and both $\Delta S$ and $\Delta Q/T$ are positive. This condition means $\frac{T}{\omega_j}<\frac{T_q}{\Omega}$, which is stronger than the classical condition. However, the rotating wave terms are dominated by $\omega_j=\Omega$ mode, so we also effectively have $T_R<T_d$ and ${\Delta Q}/{T}>\Delta S$. Similarly, if $\frac{\bar{n}_j+1}{\bar{n}_j}<\frac{1-p}{p}$, both $\Delta S$ and $\Delta Q/T$ are negative, and we can still have ${\Delta Q}/{T}>\Delta S$. If $1/2<p<1$, we must have $\frac{\bar{n}_j+1}{\bar{n}_j}>\frac{1-p}{p}$. The $\Delta Q$ is positive and $\Delta S$ is negative, so the Landauer's principle is naturally satisfied. 

From the above analyses we know that the energy flow corresponding to the rotating wave terms depends mainly on the temperature between the systems, with energy flowing from the high temperature system to the low temperature system. This is also consistent with our understanding of traditional thermodynamics.

For the counter-rotating wave terms, assuming $|I_{-,j}|^2 = 0$, as in the case of acceleration-induced transparency \cite{Kempf2022}, since there is an extra negative sign in front of the $|I_{+,j}|^2$ term in $\Delta S$, we can write it into the logarithmic function, so that we have
\begin{equation}
\Delta S = \lambda^2\sum_{j=1}^{l}\ln{\frac{p}{1-p}}[(\bar{n}_j+1)(1-p)-\bar{n}_jp]|I_{+,j}|^2, 
\end{equation}
and
\begin{equation}
\frac{\Delta Q}{T}= \lambda^2\sum_{j=1}^{l}\ln{\frac{\bar{n}_j+1}{\bar{n}_j}}[(\bar{n}_j+1)(1-p)-\bar{n}_jp]|I_{+,j}|^2.
\end{equation}
The corresponding physical processes are no longer stimulated absorption and emission, but rather a simultaneous increase or decrease in energy, so it would be meanless to compare the temperature. When $\frac{\bar{n}_j+1}{\bar{n}_j}>\frac{p}{1-p}$, the $\Delta Q$ of the quantum field theory is positive, meaning the field and qubit gain energy together. If $0<p<1/2$, increasing the energy means that the von Neumann entropy of the final state of the qubit is larger, so we have $\Delta S<0$. If $1/2<p<1$, $\Delta S$ is positive, but still less than $\Delta Q/T$. Similarly, when $\frac{\bar{n}_j+1}{\bar{n}_j}<\frac{p}{1-p}$, the $\Delta Q$ of quantum field theory is negative, implying that the field loses energy along with the qubit. In this case $p$ must satisfy $1/2<p<1$, and $\Delta S$ is negative but still less than $\Delta Q/T$. In all the cases the Landauer's principle naturally hold.

The Landauer's principle is valid for $|I_{-,j}|^2$ and $|I_{+,j}|^2$ separately, so it also holds in all cases. This is a general discussion, and since we do not qualify the motion state of the qubit, this conclusion is universal. Although the Unruh effect gives the vacuum state a ``temperature'', this temperature comes from the change in the reference system caused by the motion of the qubit, which is ultimately reflected in the counter-rotating wave terms, while the density matrix of the vacuum state does not actually change. The situation is similar in the Landauer's principle. The density matrix of the thermal field is not changed by any detector, and the motion of the qubit does not violate any of the four conditions above. It does not matter whether the qubit is in uniformly accelerated linear motion or in any other non-inertial motion. The state of motion affects the feedback of the detector, but not the validity of the Landauer's principle.

\section{Summary}
\label{section4}

In this work, we give a general description of the system evolution under the interaction between qubit and free massless scalar quantum field theory. We do not introduce restrictions on the motion state of the qubit or the quantum field state. By calculating the variation of the density matrix of the system, we give not only the reduced density matrix of the qubit, but also the energy variation of the quantum field theory itself. We analyze the different effects of the rotating and the counter-rotating wave terms on the system and prove that the motion of the qubit does not affect the validity of Landauer's principle. 

We briefly discuss some directions for future pursuit. First, since the Unruh effect has not yet been observed experimentally, one of the most important problem, even in theory, should be the enhancement of the observed effect, i.e. the contribution of the counter-rotating wave terms. A reasonable idea is given in the literature \cite{Kempf2022}, where it is proposed to compress the rotating wave terms by changing the kinematics of the qubit and to increase the value of the counter-rotating wave terms by adding the number of particles to the quantum field. We could generalize this program and look for models that can improve the observable effects. For example, in Gaussian quantum mechanics, the continuous variable method allows us to investigate the long-time evolution of the strongly coupled systems. The detector undergoes circular motion may also obtain much stronger Unruh effect \cite{Zhou:2023evt}. They may help to further amplify the contribution of counter-rotating wave terms.

Second, as mentioned earlier, the rotating wave approximation is considered a prerequisite in many models and methods of atomic physics and quantum optics. Since the counter-rotating wave term also makes an important contribution, we can reture to the previously studied topics and investigate the corrections and experimental phenomena introduced by the counter-rotating wave terms. This may bring new perspectives on atomic physics and quantum optics.

Finally, all the analyses in this work can, in principle, be set in curved spacetime and studied under different spacetime metrics, gravitational models, and quantum field theories. In this work we calculate the energy change of the quantum field theory, and in curved spacetime this leads to a perturbation in the spacetime dynamics, so studying the change in the environment can also help us better understand the impact of quantum effects on spacetime. Furthermore, in general other physical quantities of the quantum field theory can also be calculated from the change in the density matrix, but this usually requires further research because of the difficulty in the matrix diagonalization. Only in some special cases, such as when the total system is Gaussian \cite{Landi:2020bsq,Ptaszynski2019,Xu:2021ihm}, we can get some preliminary results. Since the exchange of information between the detector and the environment may also provide new perspectives on our understanding of gravity, using detectors in AdS spacetime to study the AdS/CFT correspondence may be an approach to explore \cite{Banerjee:2022dgv}. We hope to study this in our future work.

\section*{Acknowledgements}

Hao Xu thanks Yuan Sun for useful discussion and National Natural Science Foundation of China (No.12205250) for funding support.


\providecommand{\href}[2]{#2}\begingroup
\small\itemsep=0pt
\providecommand{\eprint}[2][]{\href{http://arxiv.org/abs/#2}{arXiv:#2}}

\end{document}